# Orthogonal Frequency Division Multiplexing and its Applications

Beena R. Ballal[1], Ankit Chadha[2], Neha Satam[3]

[1]Assistant Professor, Electronics and Telecommunication,
VIT, Wadala
Mumbai, India
*beenarballal@gmail.com*

[2]Student, Electronics and Telecommunication,
VIT, Wadala
Mumbai, India
*ankitrchadha@gmail.com*

[3]Student, Electronics and Telecommunication,
VIT, Wadala
Mumbai, India
*satsamsr92@yahoo.com*

**Abstract**: *Orthogonal Frequency Division Multiplexing (OFDM) is a multi-carrier modulation technique which is very much popular in new wireless networks of IEEE standard, digital television, audio broadcasting and 4G mobile communications. The main benefit of OFDM over single-carrier schemes is its ability to cope with severe channel conditions without complex equalization filters. It has improved the quality of long-distance communication by eliminating InterSymbol Interference (ISI) and improving Signal-to-Noise ratio (SNR). The main drawbacks of OFDM are its high peak to average power ratio and its sensitivity to phase noise and frequency offset. This paper gives an overview of OFDM, its applications in various systems such as IEEE 802.11a, Digital Audio Broadcasting (DAB) and Digital Broadcast Services to Handheld Devices (DVB-H) along with its advantages and disadvantages.*

## 1. Introduction

During the last few decades, growth rate of wireless technology has been accelerated to such a level that it has become ubiquitous. Progress in fiber-optics with assurance of almost limitless bandwidth and predictions of universal high-speed wireless internet access in the not-too-distant future thrive in both the popular press and technical journals [1].Wireless communication is having the fastest growth phase in history because of unprecedented evolution in the field. The kid of wireless communication is experiencing golden days due to various wireless standards such as Wi-Fi, GSM, Wimax and LTE. These standards operate within lower microwave range (2-4GHz). Due to intrinsic propagation losses at these frequencies and problem of multipath fading, it was necessary to provide a solution which can offer robustness in multipath environments and against narrowband interference and is efficient. OFDM, in all this aspects, proves to be an apt candidate by not only providing high-capacity, high-speed wireless broadband multimedia networks but also coexists with current and future systems.

Orthogonal frequency-division multiplexing (OFDM) is a method of digital modulation in which a signal is split into several narrowband channels at different frequencies. OFDM has been adopted by several technologies such as Asymmetric Digital Subscriber Line (ADSL) services, IEEE 802.11a/g, IEEE 802.16a, Digital Audio Broadcast (DAB), and digital terrestrial television broadcast: DVD in Europe, ISDB in Japan 4G, IEEE 802.11n, IEEE 802.16, and IEEE 802.20. OFDM converts a frequency-selective channel into a parallel collection of frequency flat sub channels [2].Though it is derived from frequency division multiplexing (FDM), OFDM provides many advantages over this conventional technique. In OFDM the subcarrier frequencies are chosen so that the signals are mathematically orthogonal over one OFDM symbol period. Both modulation and multiplexing are attained digitally using an inverse fast Fourier transform (IFFT) and thus, the required orthogonal signals can be generated accurately [3].

This paper is organized as follows: Section 2 describes the architecture of OFDM and Section 3 focuses on application of OFDM in various systems. Section 4 enlightens future work in this area. Finally, conclusions are presented in Section 5.

## 2. Architecture of OFDM

Practically, OFDM modulation for standard IEEE 802.20 is used by both the forward and reverse links. IEEE 802.20, also referred to as Mobile-Fi, is optimized for IP and roaming in high-speed mobile environments. This standard is ready to fully mobilize IP, opening up major new data markets beyond the more circuit-centric 2.5G and 3G cellular standards. Its main operation is to develop the specification for an efficient packet-based air interface optimized for the transport of IP-based services.

For IEEE 802.20, transmission on the forward link is divided into super frames, where each super frame consists of a preamble followed by a sequence of 25 Forward Link Physical Layer (NFLPHY) frames [4]. Transmission on the reverse link is also divided into units of super frames, with each super frame consisting of a sequence of 25 reverse link PHY frames.

In order to support cell sizes of macro, micro, and pico IEEE 802.20 should operate in a traditional cellular environment. To increase the availability of coverage area, increase throughput available to the users, and enable a higher overall spectral efficiency, advanced antenna technologies such as multi antenna at the base station should be employed.

The mathematical description for the OFDM signal is given as follows:

The low-pass equivalent OFDM signal is expressed as

$$X(t) = \sum_{k=0}^{N-1} X_k e^{j2\pi kt/T} \quad , 0 \leq t < T \quad (1)$$

this is also Discrete Fourier Transform (DFT). Here $X_k$ are data symbols which is sequence of complex numbers representing BPSK, QPSK or QAM baseband symbol, N is number of subcarriers and T is OFDM symbol line. The subcarrier spacing $\frac{1}{T}$ makes them orthogonal over each symbol period.

Sequence of OFDM symbols is given as follows:

$$S(t) = \sum_{k=-\infty}^{+\infty} X(t - kT) \quad (2)$$

To avoid ISI, a guard interval of length $T_g$ is inserted before OFDM block. During this interval, a cyclic prefix is transmitted. The signal with cyclic prefix is thus given as,

$$X(t) = \sum_{k=0}^{N-1} X_k e^{j2\pi kt/T} \quad , -T_g \leq t < T \quad (3)$$

The Fast Fourier Transform (FFT) is a computationally efficient implementation of DFT. Inverse Fast Fourier Transform (IFFT) and FFT are main modulation and demodulation techniques used in OFDM.

## 3. Application of OFDM in various systems

### 3.1 In Standard IEEE 802.11a

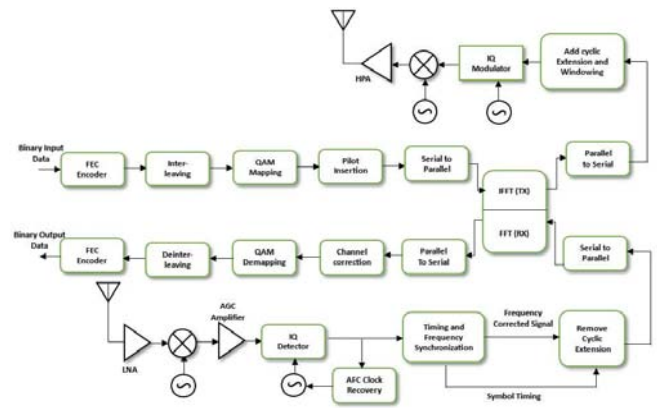

**Figure 1:** Block diagram of OFDM transceiver [6]

In the transmitter, input data which is in binary is encoded by a rate half convolution encoder. After interleaving, the binary values are converted to QAM values [5]. Four pilot values are added to each 48 data value, so that coherency at the reception point can be achieved. It gives 52 QAM values per OFDM symbol. Application of IFFT modulates the symbol onto 52 subcarriers. Cyclic prefix is added to make the system robust to multipath propagation. Narrower output spectrum is obtained by applying windowing. Using an IQ modulator, the signal is converted to analog, which is up converted to the 5 GHz band, amplified, and transmitted through the antenna.

The receiver performs the reverse operations of the transmitter, with few additional tasks. In the first step, the receiver has to estimate frequency offset and symbol timing, using special training symbols in the preamble [6]. After removing the cyclic prefix, the signal can be applied to a Fast Fourier Transform to recover the 52 QAM values of all subcarriers. The training symbols and the pilot subcarriers are used to correct for the channel response as well as remaining phase drift. The QAM values are then demapped into binary, and finally a Viterbi decoder decodes the information bits. Fig. 1 depicts block diagram of OFDM transceiver.

### 3.2 In Digital Audio Broadcasting (DAB)

Digital Audio Broadcasting (DAB) is a digital radio technology for broadcasting radio stations, used in several countries, especially in Europe. It has 4 transmission modes with different parameters as shown in the Table 1.

**Table 1:** Digital Audio Broadcasting parameters [7]

| Parameters | Mode I | Mode II | Mode III | Mode IV |
|---|---|---|---|---|
| No. of sub-carriers | 1536 | 384 | 192 | 768 |
| Sub-carrier spacing | 1kHz | 4kHz | 8kHz | 2kHz |
| Symbol time | 1.246ms | 311.5us | 155.8us | 623us |
| Guard time | 246us | 61.5us | 30.8us | 123us |
| Carrier frequency | <375MHz | <1.5GHz | <3GHz | <1.5GHz |
| Transmitter separation | <96km | <24km | <12km | <48km |

The DAB transmitted data consists of number of signals sampled at a rate of 48 kHz with a 22-bit resolution [7]. This signal is then compressed at rates ranging from 32 to 384 kbps, depending upon the desired quality. The resulting digital data is then divided into frames of 24 ms. DAB uses differential QPSK modulation for the sub-carriers. A null-symbol indicates the start of the frame. A reference OFDM symbol is then sent to serve as a starting point for the differential decoding of the QPSK subcarriers. Differential Modulation avoids the use of complicated phase-recovery schemes. DAB uses a rate quarter convolutional code with a constraint length of 7 for error-correction. Interleaving is used to separate the coded bits in the frequency domain as much as possible, which avoids large error bursts in the case of deep fades affecting a group of sub-carriers.

### 3.3 In DVB-H: Digital Broadcast Services to Handheld Devices

Digital Video Broadcasting (DVB) is a set of internationally accepted standards for digital television. DVB-H is one of the established mobile TV formats. It permits transmission of very large files and can operate on 5, 6, 7 or 8 MHz bandwidth.

DVB-H uses OFDM air interface technology, and includes a technique for power reduction in the tuner. It uses time-slicing so that the tuner can be switched off most of the time and is only on during short transmission bursts. This allows the tuner to operate over a reduced input bandwidth and also conserves power. OFDM is a very good choice for a mobile TV air interface. It offers good spectral efficiency, immunity to multi-path, good mobile performance, and it works well in single-frequency networks such as those planned for mobile TV.

The structure of DVB-H is depicted in fig. 2.

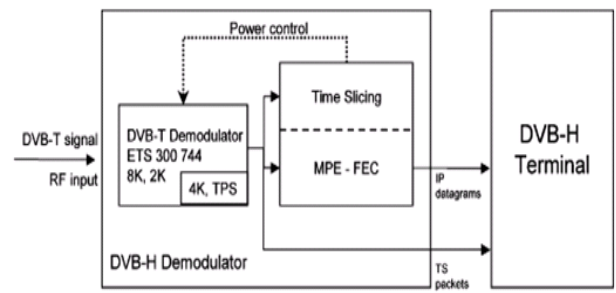

**Figure 2:** Conceptual structure of a DVB-H receiver [8]

It includes a DVB-H receiver (a DVB-T demodulator, a time-slicing module, and an optional MPE-FEC module) and a DVB-H terminal. The DVB-T demodulator recovers the MPEG-2 transport stream (TS) packets from the received DVB-T RF signal [8]. It offers three transmission modes: 8K, 4K, and 2K with the corresponding signaling. The time-slicing module controls the receiver to decode the wanted service and shut off during the other service bits. It aims to reduce receiver power consumption while also enabling a smooth and seamless frequency handover. The MPE-FEC module, provided by DVB-H, offers in addition to the error correction in the physical layer transmission, a complementary FEC function that allows the receiver to cope with particularly difficult reception situations.

The advantages of DVB-H are as follows [9]:

- Carriers - In DVB-H, carriers can use any additional spectrum that they might own for DVB-H broadcasting and be an infrastructure player.
- Spectrum Availability - In U.S., DVB-H will be organized using clear and "ready-for-use" spectrum available today, without interfering with existing analog TV stations or other TV or wireless services.

### 4. Advantages and disadvantages

Advantages of OFDM are listed as follows:

- OFDM makes resourceful utilization of the spectrum by overlapping. By dividing the channel into narrowband flat fading sub channels, OFDM is more resistant to frequency selective fading than single carrier systems.
- It can easily adapt to severe channel conditions without complex time-domain equalization.
- It reduces ISI and IFI through use of a cyclic prefix and fading caused by multipath propagation.
- Using sufficient channel coding and interleaving lost symbols can be recovered.
- Channel equalization becomes simpler than by using adaptive equalization techniques with single carrier systems.
- OFDM is computationally capable by using FFT techniques to implement the modulation and demodulation functions.
- It is less sensitive to sample timing offsets than single carrier systems are.
- It is robust against narrow-band co-channel interference.
- Unlike conventional FDM, tuned sub-channel receiver filters are not required.
- It facilitates single frequency networks (SFNs); i.e.,

transmitter macro diversity.

The disadvantages are as follows:

- The OFDM signal has a noise like amplitude with a very large dynamic range; hence it requires RF power amplifiers with a high peak to average power ratio.
- It is more sensitive to carrier frequency offset and drift than single carrier systems are due to leakage of the DFT.
- It is sensitive to Doppler shift.
- It requires linear transmitter circuitry, which suffers from poor power efficiency.
- It suffers loss of efficiency caused by cyclic prefix.

## 5. Conclusion

OFDM has promising future in wireless networks and mobile communications. Growth in number of worldwide customers for wireless networks and ever-increasing demand for large bandwidth has given birth to this technology. OFDM is already playing an important role in WLAN and will be part of MAN too. In coming years, it will surely dominate the communication industry. Also, Wimax and 802.20 use OFDM-MIMO, which is emerging as the main technology for future cellular packet data networks, including 3GPP long-term evolution and 3GPP2 air interface evolution as well. Although OFDM has proven itself with packet-based data, it is not yet clear whether the technology can either handle large numbers of voice customers or work with voice and data as well as CDMA.

**Author Profile**

**Beena Ballal** is an Assistant Professor at Vidyalankar Institute of Technology; Mumbai. She has a teaching experience of 8 years and has interest in Optical fibers and Digital logics. She has two national and three international publications to her credit.

**Ankit Chadha** is currently pursuing his undergraduate degree in Electronics and Telecommunication Engineering discipline at Vidyalankar Institute of Technology, Mumbai. His fields of interest include Image Processing, Embedded Systems.

**Neha Satam** is currently pursuing his undergraduate degree in Electronics and Telecommunication Engineering discipline at Vidyalankar Institute of Technology, Mumbai. Her fields of interest include Image Processing, Wireless Communication.